\documentclass[useAMS,usegraphicx,usenatbib]{mn2e}
\usepackage{amsfonts,amssymb,amsmath}
\usepackage{xcolor}
\usepackage{graphicx}
\usepackage{soul}
\usepackage{epstopdf}
\usepackage{ulem}
\usepackage{multicol}
\title[Mass-richness relation using the SDSS database.]{Weak lensing measurement of the mass-richness relation using the SDSS database.}
\author[Gonzalez et al.]{Elizabeth Johana Gonzalez$^{1,2,3}$\thanks{E-mail:
elizabethjgonzalez@oac.unc.edu.ar}, 
Facundo Rodriguez $^{1,2}$, 
Diego Garc\'{\i}a Lambas$^{1,3}$,
\newauthor 
Manuel Merch\'an$^{1,3}$,
Gael Fo\"{e}x $^{4}$,
Mart\'in Chalela$^{1,2}$\\
$^{1}$ Instituto de Astronom\'{\i}a Te\'orica y Experimental, (IATE-CONICET),
 Laprida 854, X5000BGR, C\'ordoba, Argentina.\\
$^{2}$ Facultad de Matem\'atica, Astronom\'ia y F\'isica, FAMAF, Universidad Nacional de C\'ordoba, X5000BGR, C\'ordoba, Argentina.\\
$^{3}$ Observatorio Astron\'omico de C\'ordoba, Universidad Nacional de C\'ordoba, Laprida 854, X5000BGR, C\'ordoba, Argentina.\\
$^{4}$ Max Planck Institute for Extraterrestrial Physics, Giessenbachstrasse, 85748 Garching}

\begin{document}


\pagerange{\pageref{firstpage}--\pageref{lastpage}} \pubyear{2015}

\maketitle

\label{firstpage}

\begin{abstract}

We study the mass-richness relation using galaxy catalogues and images from the \textit{Sloan Digital Sky Survey}. We use two independent methods, in the first one, we calibrate the scaling relation with weak-lensing mass estimates. In the second procedure we  apply a background subtraction technique to derive the probability distribution, $P(M \mid N)$, that groups with $N$-members have a virialized halo mass $M$. Lensing masses are derived in different richness bins for two galaxy systems sets: the maxBCG catalogue and a catalogue based on a group finder algorithm developed by \citeauthor{Yang05}. MaxBCG results are used to test the lensing methodology. The lensing mass-richness relation for the \citeauthor{Yang05} group sample shows a good agreement with $P(M \mid N)$ obtained independently with a straightforward procedure.

\end{abstract}


\section{INTRODUCTION}

Within the current cosmological paradigm, the Universe is dominated by the presence of still unidentified, weakly interacting particles, the so-called cold dark matter model \citep[CDM; eg. ][]{Bond82, Peebles82, Bertone04, Komatsu09, Garrett11, Peter12}. In this scenario, structure formation takes place driven by gravitation collapse of initial density fluctuations leading to  localized, highly overdense clumps of  dark matter, dubbed as halos. 

The relation between galaxies and dark matter halos can provide relevant information regarding several aspects of the matter distribution in the Universe through observations. There are several methods to study how galaxies populate dark matter halos \citep{Guo10}.  One way is to follow galaxy formation in $N$-body simulations combined with hydrodynamical \citep[e.g.][]{Cen00, Springel03, Keres05, Sijacki07} or semi-analytical models \citep[eg.][]{Kauffmann99, Springel01, Hatton03,Springel05, Kang05} to consider baryonic evolution. These methods provide information of diverse galaxy properties as a function of time and have been successful in reproducing observations. Nevertheless there are several ad-hoc parameters in the recipes employed to model star formation, blackhole evolution, and, mostly important, their associated feedback processes. An alternative method to study how galaxies populate halos, is by linking galaxies to subhalos/halos, according to the relation between galaxy luminosity functions and halo mass functions, assuming a unique and monotonic relation between these functions \citep[e.g.][]{Vale04,Conroy06,Shankar06,Conroy07,Baldry08,Moster10}. Also, simple statistical models such as the Halo Occupation Distribution (HOD) can be used to link galaxies with halos, irrespective of their physical properties. HOD describes the probability distribution, $P(N \mid M_{h})$, that a virialized halo of mass $M_{h}$ contains $N$ galaxies \citep[e.g.][]{Jing98a, Jing98b, Ma00, Peacock00, Seljak00, Scoccimarro01, Berlind02, Cooray02, Bosch03, Berlind03, Zheng05, Yang08}. This leads to the occupation number of a given halo, namely, the number of galaxies above a given luminosity or stellar mass threshold, as a function of the halo mass. Constraining this relation is important to test cosmological, semi-analytical, galaxy formation and evolution models. HOD has been mainly determined by assuming a functional form and fitting the free parameters using statistical data of galaxy  abundance and clustering. A direct measurement of the HOD requires a determination of the number of galaxies of the group and the total mass of the halo. However, estimating accurate halo masses is a great challenge, in particular for poor groups, given the considerable uncertainties of dynamical masses and their low X-ray emission.

Weak lensing has proved to be an excellent technique for mass determination given that it is sensible to both, barionic and non-barionic matter. Nevertheless, detecting weak lensing signal is a hard task since the small shape distortions that need to be measured are strongly affected by the atmosphere and the instruments. Therefore, this technique has been mostly applied to galaxy clusters where the mass density is high enough to obtain a reliable signal. In order to apply weak lensing techniques to low-mass galaxy systems, such as poor clusters and groups ($\sim 10^{13} M_{\odot} $), stacking techniques are a powerful tool to increase the signal-to-noise ratio (S/N) and thus, to derive reliable measurementes of the composite (average) galaxy system \citep{Leauthaud10,Melchior13,Rykoff08,Foex14}. Furthermore, a weak lensing analysis allows  to study mass profiles at large distances from the lens centre. This is due to light bundles from distant background galaxies at different angular distances from the lensing system which provide information beyond the luminous extent of the galaxy system. 

In previous works, the lensing mass - richness relation has not taken into account HOD modeling \citep{Becker07,Johnston07,Reyes08,Rykoff08,Mandelbaum08,Rozo09, Hilbert10, Foex12, Uitert16}. In this work, we obtain the mass-richness relation using two independent approaches: HOD and lensing mass - richness relation. With this aim, we determine the $P(M_h \mid N)$ following \citet{Rodriguez15} using a background substraction technique. We derive lensing masses for a sample of galaxy groups in order to compare two independent, compatible relations, and to set the basis for future projects. 

This work is organized as follows: In section \ref{sec:theory} we link the HOD with weak lensing mass-richness relation. Section \ref{sec:data} describes the sample of galaxy clusters and groups studied. Details of the lensing analysis are provided in section \ref{sec:method} and in section \ref{sec:results} we discuss our results, compare them with previous works and discuss the mass-richness relation. Finally, in section \ref{sec:test},  we summarize our results. We adopt a standard cosmological model $H_{0}$\,=\,70\,km\,s$^{-1}$ \,Mpc$^{-1}$, $\Omega_{m}\,=\,0.3$, and $\Omega_{\Lambda}\,=\,0.7$.

\section{MASS-RICHNESS RELATION AND HOD}

\label{sec:theory}

Masses of galaxy groups and clusters  can be estimated using different observables such as X-ray emission, weak lensing shear, spectroscopic information, cluster richness, etc. The study of the relations linking the total mass with other physical quantities is important since they allow to derive system masses from simple observables. A well studied relation is the weak lensing mass and the optical richness ($M_{Lens} - N$) which presents a logarithmic slope close to one, in agreement with the simplest model of structure formation \citep{Kravtsov04}.

HOD and $M_{Lens} - N$ relation are two idependent descriptions of the dark matter content  in galaxy clusters and they have not been treated together in previous works. The HOD links halo mass with the number of galaxies, connecting systems of a given mass with the average number of galaxy members, $P(N \mid M_{h})$. On the other hand, weak lensing stacking techniques allow the computation of the average halo mass for a sample of groups with a given richness, $P(M_h \mid N)$. Thus, to compare both relations we have estimated the $P(M_h|N)$ using the same technique as for the HOD (i.e. $P(N|M_h)$).

In order to obtain $P(M_h|N)$ we use a background subtraction procedure, as described by \citet{Rodriguez15}. This technique involves counting objects in a region where there is a known signal superimposed on uncorrelated noise, which is subtracted by using a statistical estimation. The signal corresponds to the overdensities associated to the galaxy groups, while the noise is associated to foreground and background galaxies (interlopers). Following this procedure, we combine galaxy systems from spectroscopic surveys with catalogues without redshift information, which makes it possible to estimate the $P(M_h|N)$ in a wider range of magnitudes and, at the same time, statistics is improved. Absolute magnitudes are computed assuming that all galaxies are located at the group mean redshift. Then, we count galaxies within a circle of a group projected characteristic radius centred on each group, with absolute magnitudes $M\leq M_{lim}$. To estimate the noise it is assumed that the galaxy distribution is close to uniform in the large-scale average, while groups are local overdensities. As it is not possible to determine straightforward the interlopers number, a statistical method is required. Taking into account the hierarchical behaviour of the large scale structure, it is known that a given overdensity is always immersed in a larger structure, therefore the background contribution is computed by counting the number of galaxies that meet the selection criteria within an annulus centred on each galaxy group. Finally, the HOD can be estimated by subtracting the local background density multiplied by the projected area for each group (full details and tests of this procedure are given by \citet{Rodriguez15}). 

We determine lensing masses using stacking techniques for two samples of galaxy clusters/groups, the maxBCG cluster catalogue \citep{Koester07b} and a group sample from \citet{Yang12}. MaxBCG sample has been extensively analysed using gravitational lensing \citep{Johnston07, Sheldon09, Mandelbaum08, Tinker12}. We use this sample to test our lensing analysis implementation, which in spite of considerable simplifications, gives a very good agreement with previous works. Then, we apply our lensing analysis to the \citet{Yang12} galaxy group sample, in order to compare it with our $P(M_h|N)$ relation.


\section{SAMPLES	 AND DATA ACQUISITION}

\label{sec:data}

\subsection{Data acquisition}

The Sloan Digital Sky Survey \citep[SDSS,][]{York00} is the largest photometric and spectroscopic survey at present. It was constructed using a 2.5\,m telescope at Apache Point Observatory in New Mexico. The seventh data release \citep[SDSS-DR7,][]{Abazajian09} comprises 11663 square degrees of sky imaged in five wave-bands (\textit{u, g, r, i} and \textit{z}) containing photometric parameters of 357 million objects. The spectroscopic survey is a magnitude limited sample to $r_{lim} < 17.77$ (Petrosian magnitude), most of galaxies span a redshift range $0 < z < 0.25$ with a median redshift of 0.1 \citep{Strauss02}. The SDSS observing mode is time-delay-and-integrate, with the camera reading out at the scan rate, resulting in an effective exposure time of 54 seconds. Each image is 10 by 13 arcminutes, corresponding to 2048 by 1489 pixels, with a pixel size of 0.396". 

In order to compute the $P(M_h|N)$ we use photometric and spectroscopic data from SDSS-DR7 as in \citet{Rodriguez15}. Images for the weak lensing analysis are obtained from data release 10 (SDSS-DR10, http://data.sdss3.org/sas/dr10) in $r$ and $i$ bands. Data Release 10 includes all prior SDSS imaging data, which allows us to select the image in the field of a given galaxy group detected in DR7, with seeing conditions good enough to perform the lensing study.  For the sake of simplicity, we analyse only one image for each lensing system. We determine the seeing of the $i$-band frames, ranked according to the centres frames distances to the lens. This process stops when the seeing is  0.9" or lower, or when the centre of the lensing system is out of the field-of-view of the image (excluding an edge of 50 pixels per side). The frame with the lowest seeing value is used for the analysis, discarding those lensing systems with  values greater than 1.3".

Photometry is performed in both bands and shape measurements are done in band $i$ since it has better seeing conditions. Details regarding detection, photometry and classification of the sources, as well as shape measurements are given in \citet{Gonzalez15}. For the lensing analysis we only consider galaxies brighter than $m_{r} = 21.0$ (where $m_{r}$ is the measured apparent magnitude in $r$ band corrected by galactic extinction, computed following \citet{Schlegel98} at the position of each lensing group). We also restrict the objects to those with a good pixel sampling by using only galaxies with FWHM $>$ 5 pixels. That also ensures that the shape measurement is less affected by the point spread function (PSF), given that the mean seeing is $\sim 1.0 " = 2.5$\,pixels.

\subsection{MaxBCG Catalogue}

We used the galaxy cluster catalogue \citep{Koester07b} constructed employing the maxBCG red-sequence method \citep{Koester07a} from the SDSS photometric data. This method is based on three primary features of galaxy clusters: 1) high galaxy density contrast, 2) brightest members share similar colours and 3) presence of a brightest galaxy member (BCG) that is usually at rest located at the cluster's centre of mass \citep{Oegerle01}.

The first step for the maxBCG algorithm is to compute for each galaxy two independent likelihoods. The first one, is the likelihood that a galaxy is spatially located in an overdensity of E/S0 galaxies with similar \textit{g-r} and \textit{i-r} colours, and the second one is the likelihood that it might be a BCG according to its colour and magnitude. The redshift that maximizes the product of these likelihoods is adopted for each galaxy and constitutes a first estimate for the cluster redshift. After that, each galaxy is treated as a potential BCG, and for the clusters associated, a list of members is constructed. The cluster characteristic size, $R_{200}$, is defined as the radius in which the density of galaxies with -24 $\leq M_{r} \leq$ -16 is 200 times their mean number density. $R_{200}$ is estimated based on an the richness-size relation determined by \citet{Hansen05} using an initial guess for the cluster richness ($N_{gal}$). In turn, $N_{gal}$ is obtained by counting the number of galaxies brighter than 0.4\,$L_{\ast}$ within 1$h^{-1}$Mpc of this potential centre. Also, these galaxies are required to be fainter than the BCG candidate and to have colors matching the E/S0 ridgeline. The potential BCGs are ranked by decreasing maximum likelihood, and the first object in the list becomes the first cluster centre. All remaining objects in the list within a redshift range, $z \pm 0.02$, and within the radius $R_{200}$, are discarded as BCGs candidates. The process is repeated for the next object  and after cycle through the list, remaining galaxies are taken as the BCGs of the final cluster list. 

The final catalogue contains 13,823 galaxy clusters and includes measured properties such as location, redshift (photometric, and spectroscopic when available), and several richness and mass estimators. For our analysis we use celestial coordinates, redshifts and $N_{200}$ (defined as the number of E/S0 ridgeline members brighter than 0.4\,$L_{\ast}$ within $R_{200}$ of the cluster centre). The purity and completeness of this catalog are above 90$\%$ for $N_{200} \geq 10$ across $0.1 < z < 0.3$.\\

\subsection{Yang group sample}
\label{yang_sample}

We use a sample of the SDSS galaxy group catalogue of \citet{Yang07}, constructed using the adaptive halo-based group finder presented in \citet{Yang05}, but updated to DR7 \citep{Yang12}. This group finder uses conventional friends-of-friends (FOF) algorithm combined with the properties of the halo population. It uses a FOF algorithm to assign galaxies to groups. Geometrical centres of all FOF groups with more than two galaxies are considered as potential centres of groups. Galaxies that were not linked to a FOF group but found to be the brightest galaxy in a cylinder of radius 1$h^{-1}$Mpc and velocity depth $\pm$500\,km\,s$^{-1}$, were also considered as potential groups centres. Once the potential centres are obtained, a total luminosity is computed for each group and the mass is estimated using a model for the mass-to-light ratio. This mass is used to estimate the size and velocity dispersion of the underlying halo that hosts the group, which in turn is used to determine the galaxy group members in redshift space. The procedure is repeated until convergence. This method has the advantage of identifying galaxy groups with only one member detected. 

In this work we analyse galaxy groups that have at least one member with $r$-band absolute magnitude,  $M_{r} < -21.5$. For groups with $2 \leq N_{member} \leq 6$  we use objects ranging from z=0.1 to z=0.2.  For the sample with $N_{member} = 1$ we use a narrower range of redshifts, $0.1 <  z < 0.15$. This is due to the great amount of time consumed by the lensing analysis, so in order to decrease the computing time, we reduce the number of systems.  The resulting sample of analysed objects contains 18208 groups.

\section{WEAK LENSING ANALYSIS}
\label{sec:method}
\subsection{Stacking technique}

SDSS gives access to a large sky coverage and image data. Nevertheless, given the short exposure time (53.9 seconds for each pixel), images are not deep enough to perform a weak lensing analysis of individual objects. Furthermore, in this work we analyse galaxy systems with masses $\sim 10^{13} M_{\odot} $ which are expected to have a low weak lensing signal. This, in turn, leads to a low value of the shape distortion of source galaxies which is related to the shear components, $\gamma$, and carries the information of the lens gravitational potential -- e.g. for a source at $z=0.3$ and a lens mass $\sim 10^{13} M_{\odot} $ at $z=0.1$, gives $\gamma \sim 0.01$ at 100\,kpc, significantly lower than the main source of noise, i.e. the dispersion of the intrinsic galaxy ellipticity distribution ($\sim 0.2 - 0.3$).
 
To overcome this problem we use stacking techniques which consist in combining several systems to derive the average mass. Since the noise scales as $1/\sqrt{N}$, where $N$ is the number of sources, the use of stacking techniques can provide a lensing signal with suitable confidence level. Furthermore, it reduces the impact of substructures present in the individual systems and their deviations from sphericity. Finally, when we average the signal of many lenses, the effects produced by the large scale structure are averaged out producing only an additional statistical noise. Taking into account the mentioned advantages of the stacking techniques, we combine several subsamples of galaxy groups and clusters, according to richness. The procedure is carried out following the formalism given by \citet{Foex14}. 

To estimate the tangential shear component, $\tilde{\gamma}_{T}$, we use the ellipticity components of background galaxies \citep[see ][for details about galaxy selection]{Gonzalez15}, $\tilde{\gamma}_{T,j}(r) = \langle e_{T} \rangle_{j}$, where $\langle e_{T} \rangle$  is the average tangential ellipticity component of the $N_{Sources,j}$ galaxies, located at a radius $r \pm \delta r$ from the $j$th lens. The average on annular bins of the ellipticity component tilted at $\pi/4$, $\langle e_{X} \rangle_{j}$, should be zero and correspons to the cross shear component, $\tilde{\gamma}_{X,j}(r)$.

The average mass density contrast of $N_{Lens}$ circular-symmetric lenses is computed according to the tangential ellipticity component, $e_{T,ij}$, of each source $i$ corresponding to the lens system $j$, according to:
\begin{large}
\begin{equation}
\label{eq:sigma}
\langle \Delta \tilde{\Sigma}(r) \rangle = \frac{\sum_{j=1}^{N_{Lens}} \sum_{i=1}^{N_{Sources,j}} \omega_{ij} \times e_{T,ij} \times \Sigma_{crit,j}}{\sum_{j=1}^{N_{Lens}} \sum_{i=1}^{N_{Sources,j}} \omega_{ij}}
\end{equation}
\end{large}\\
where  $\omega_{ij}$ are the weights considered for each source galaxy and $\Sigma_{crit,j}$ is the critical density for all the sources of the lens $j$, defined as:
\begin{equation*}
\Sigma_{crit,j} = \dfrac{c^{2}}{4 \pi G} \dfrac{1}{\langle \beta_{j} \rangle D_{OL_{j}} }
\end{equation*}
here, $D_{OL_{j}}$ is the angular diameter distance from the observer to the $j$th lens, $G$ is the gravitational constant, $c$ is the light velocity and $\langle \beta_{j} \rangle$ is the geometrical factor defined as the average ratio between the angular diameter distance from the galaxy source $i$ to the lensing system $j$, $D_{LS_{j}}$, and the angular diameter distance between the observer and the source, $D_{OS_{i}}$ ($\langle \beta_{j} \rangle = \langle D_{LS_{j}}/D_{OS_{i}} \rangle_{i}$). $\Sigma_{crit,j}$ is estimated for each lensing group using a catalogue of photometric redshifts as described in \citet{Gonzalez15}.

Since shape parameters are estimated using Markov-Chain Monte Carlo sampling \citep[details regarding the code employed for the shape measurements are given in ][]{Bridle02}, each galaxy is measured twice and the difference between the first and second measurement of the ellipticity is taken as the shape measurement error, $\sigma_{SE,i}$. Only the galaxies with $\sigma_{SE,i}$ lower than 0.1 are kept for the analysis. We weight the ellipticities according to the adopted error and the scaled size of the source galaxy:
\begin{large}
\begin{equation*}
\omega_{ij}=\dfrac{1}{(R_{ij}^{2} + \sigma_{SE,i}^{2} )\times \Sigma_{crit,j}^{2}}
\end{equation*}
\end{large}\\
where $R_{ij}$ is the scaled Gaussian full width at half-maximum of the source galaxy (FWHM$_{i}$) to the maximum FWHM$_{i}$ of the image $j$, $R_{ij} = $FWHM$^{max}_{j}/$FWHM$_{i}$. 

The uncertainties associated to the estimator $\langle \Delta \tilde{\Sigma}(r) \rangle$ are computed taking into account the noise due to the galaxies intrinsic ellipticity, $\sigma_{\gamma} \approx 0.25$,

\begin{large}
\begin{equation} 
\label{eq:err}
\sigma^{2}_{\Delta \tilde{\Sigma}}(r)=\dfrac{\sum_{j=1}^{N_{Lens}} \sum_{i=1}^{N_{Sources,j}}( \omega_{j} \times \sigma_{\gamma} \times \Sigma_{crit,j})^{2}}{\left(\sum_{j=1}^{N_{Lens}} \sum_{i=1}^{N_{Sources,j}}  \omega_{j}\right)^2}
\end{equation}
\end{large}

Finally, we compute the total S/N as follows:
\begin{large}
\begin{equation} 
\label{SN}
\left(\frac{S}{N}\right)^{2}=\sum_{i} \dfrac{ \langle \Delta \tilde{\Sigma}(r_{i})\rangle ^{2}}{\sigma^{2}_{\Delta \tilde{\Sigma}}(r_{i})}
\end{equation}
\end{large}\\
where the sums run over all the bins in radius used to fit the profile.

Since redshift information is not available for all galaxies in our sample, there can be a residual contamination by faint group members. These galaxies weaken the lensing signal, since they are not sheared. Consequently, a smaller shear can be measured, and this derives in a lower galaxy system mass. To overcome this problem, we follow the method proposed by \citet{Hoekstra07} according to which the observed shear is multiplied by a factor $1+f_{cg}(r)$, where $f_{cg}(r)$ is the fraction of galaxy members that remain in the catalogue of background galaxies. To estimate $f_{cg}(r)$ we fit a $1/r$ profile to the galaxy excess relative to the background level and we correct the measured shear according to the distance to the lensing system centre. It has been noticed that a $1/r$ profile could lead to an overestimation of the contamination in the central part of galaxy clusters ($r < 500\,h^{-1}_{70}$\,kpc), since some rich systems may present a central core \citep{Hoekstra15}. Nevertheless, given that most of the analysed objects are low mass  systems, we do not consider here, the presence of a central core.

\subsection{Mass profile of stacked galaxy groups}

The mass density contrast profiles obtained from Equation\,\ref{eq:sigma} can be used to estimate lensing masses by fitting a parametrized physical model. This usually comprises three components: the central stellar mass contained in the BCG, the group/cluster main dark matter halo, and the contribution from other neighboring mass concentrations \citep[eg.,][]{Mandelbaum05,Johnston07,Leauthaud10,Oguri11,Umetsu14}. The first component has a significant influence on small scales (up to $\sim50$\,kpc), while the third halo component has a dominant contribution well beyond the virial radius of the main halo \citep{Oguri11}. 

Profiles obtained from the stacked weak-lensing analysis are built assuming the centre of the lensing system as the position of the brightest galaxy of each galaxy group/cluster (BCG). As described in Section\,\ref{sec:data}, we use one image for each lens and, taking into account the limited angular size of SDSS frames, we do not consider the third halo component to model the mass profile. Also, we avoid fitting the central parts in order to use only one simple model that describes the main component of the group/cluster dark matter halo. We compute the profiles beyond $90\,h_{70}^{-1}$\,kpc, where the signal becomes significantly positive, to avoid the regions in which the BCG gravitational potential is dominant. 

Average density contrast profiles are constructed using non-overlapping concentric logarithmic annuli. Since the results do not show a strong dependence on annuli sizes, we have adopted its value in order to obtain the lowest profile fit errors. 

Two mass models are used to fit the density profile: a singular isothermal sphere (SIS) and a NFW profile \citep{Navarro97}. The SIS profile is the simplest density model for describing a relaxed massive sphere with a constant value for the isotropic one dimensional velocity dispersion, $\sigma_V$. Dynamical studies of galaxies are consistent with a mass profile following approximately an isothermal law \citep[eg.,][]{Sofue01}. The shear ($\gamma_{\theta}$) and the convergence ($\kappa_{\theta}$) at an angular distance $\theta$ from the lensing system centre, scaled for a source at $z \rightarrow \infty$, are directly related to $\sigma_V$ by
\begin{equation}
\kappa_{\theta} = \gamma_{\theta} = \dfrac{\theta_{E}}{2 \theta} \frac{1}{\langle \beta \rangle}
\end{equation}
where $\theta_{E}$ is the critical Einstein radius defined as:
\begin{equation}
\theta_{E} = \dfrac{4 \pi \sigma_{V}^{2}}{c^{2}} \frac{D_{LS}}{D_{OS}}
\end{equation}

From this model we can compute the $M_{200}$ mass defined as \mbox{$M_{200}=200\rho_{crit}(z)\dfrac{4}{3}\pi\,R_{200}^{3}$}, where $R_{200}$ is the radius that encloses a mean density equal to 200 times the critical density ($\rho_{crit} \equiv 3 H^{2}(z)/8 \pi G$; $H(z)$ is the redshift dependent Hubble parameter and $G$ is the gravitational constant), as \citep{Leonard10}:
\begin{equation}\label{eq:MSIS}
M_{200} =  \dfrac{2 \sigma_{V}^{3} }{\sqrt{50} G H(z)} 
\end{equation} 

Alternatively, we use the NFW profile that is derived by fitting the halo density profile
in numerical simulations of cold dark matter halos \citep{Navarro97}. This profile depends on two parameters, the virial radius, $R_{200}$, and a dimensionless concentration parameter, $c_{200}$. The density profile follows

\begin{equation*}
\rho(r) =  \dfrac{\rho_{crit} \delta_{c}}{(r/r_{s})(1+r/r_{s})^{2}} 
\end{equation*}
where $r_{s}$ is the scale radius, $r_{s} = R_{200}/c_{200}$ and $\delta_{c}$ is the cha\-rac\-te\-ris\-tic overdensity of the halo

\begin{equation*}
\delta_{c} = \frac{200}{3} \dfrac{c_{200}^{3}}{\ln(1+c_{200})-c_{200}/(1+c_{200})}  
\end{equation*}

We use the lensing formulae for the spherical NFW density profile from \citet{Wright00}. There is a well-known degeneracy between the parameters $R_{200}$ and $c_{200}$ when fitting the shear profile in the weak lensing regime. This is due to the lack of information on the mass distribution near the cluster centre, and only a combination of strong and weak lensing can break this degeneracy and provide useful constraints on the concentration parameter. To overcome this problem, we decide to fix the concentration parameter using the relation $c_{200}(M_{200},z)$ given by \citet{Duffy11}. We use the $M_{200}$ mass estimates from the SIS model  (Equation\,\ref{eq:MSIS}) and the weight average redshift of stacked lenses according to the number of background galaxies of each lens, $\langle z_{Lens} \rangle$. The particular choice of the $M_{200}-c_{200}$ relation has not a significant impact on the final mass values, with uncertainties dominated by the noise of the shear profiles. Thus, once $c_{200}$ is fixed, we fit the profile with only one free parameter: $R_{200}$.

To derive the parameters of each mass model profile we perform a standard $\chi^{2}$ minimization:
\begin{large}
\begin{equation}
\chi^{2} = \sum^{N}_{i} \dfrac{(\langle \tilde{\Sigma}(r_{i})  \rangle - \tilde{\Sigma}(r_{i},p))^{2}}{\sigma^{2}_{\Delta \tilde{\Sigma}}(r_{i})}
\end{equation}
\end{large}\\
where the sum runs over the $N$ radial bins of the profile and the model prediction. $p$ refers to either $\sigma_{V}$ for the SIS profile, or $R_{200}$ in the case of the NFW model. 
Errors in the best-fitting parameters are computed according to the variance of the parameter estimate.

\subsection{Systematic errors in mass determinations}
\label{susec:syserrores}
In this section we discuss the uncertainties related to miscentring problems, redshift estimation of background galaxies and sample dispersion. We do not take into account errors regarding background sky obscuration \citep{Simet14} given that this effect is negligible for SDSS. 

Centring the profile on the brightest galaxy assumes that it is correctly identified and that it is actually the centre of the gravitational potential. BCG offsets from the system gravitational potential centre  could significantly suppress the lensing signal in the inner parts, leading to mass underestimations \citep{Johnston07, Mandelbaum08}. \citet{Uitert16} found that only $\sim\,30\%$ of the clusters they analysed had the BCG located at the centre of the halo; the remaining BCGs followed a 2D-Gaussian distribution, whose width, $\sigma_s$, ranges from 0.2 up to 0.4\,$h_{70}^{-1}$\,Mpc for the most massive systems ($ > 5 \times 10^{13}h_{70}^{-1}M_{\odot}$). They also found that the ratio $\sigma_s/R_{200}$ remains constant at $0.44 \pm 0.01$. In order to test how miscentring affects our lensing mass determinations, we fit 500 profiles computed according to a random centre, generated following a 2D gaussian distribution centred on the BCG and with a dispersion value of $\sigma_s = 0.44 \times R_{200}$, using the estimated $R_{200}$ radius. The fitted parameters show a gaussian distribution with mean values $\sim 3 \%$ lower than those derived using profiles centred on the BCG. This systematic difference was taken into account in the final measured parameters.

Our catalogue does not have enough redshift information to directly estimate the geometrical factor $\beta$, and the limiting magnitude to consider that a galaxy is behind the lens system \citep[this is described in ][]{Gonzalez15}. Therefore, we use the catalogue of photometric redshifts computed by \citet{Coupon09}, based on the public release Deep Field 1 of the Canada-France-Hawaii Telescope Legacy Survey, which is complete down to $m_{r} = 26$. After applying the same photometric cuts as for selecting background galaxies and taking into account the apropiate magnitude transformations we obtain $\langle \beta \rangle$. This value is fairly insensitive to the detailed redshift distribution, as long as the mean source redshift is substantially larger than the lens redshift   \citep[$z<0.3$][]{libro}, which is the case in our sample. In order to consider the contamination by foreground galaxies with our selection criteria, we set $\beta(z_{phot} < z_{lens}) = 0$  which outbalances the dilution of the shear signal by these unlensed galaxies. Deep Field\,1 covers a sky region of 1\,degree$^2$, thus to estimate the cosmic variance, we divide the field in 25 non-overlapping areas of $\sim$\,144\,arcmin$^2$ and we compute $\langle \beta \rangle$ at $z=0.18$ and $z=0.14$ for each area (these are the average redshifts of the maxBCG and Yang group samples, respectively). The uncertainties in $\langle \beta \rangle$ due to cosmic variance are estimated according to the scatter among the values for each area, obtaining  $\sim 0.04$ and $\sim 0.05$, respectively. Errors in $\langle \beta \rangle$ are lower than $7\%$, which represents an uncertainty of $\sim\,9\%$ in mass. These uncertainties were taken into account in the error estimation of the fitted parameters, and propagated to the resulting system masses.

 In order to test the stability of the results of each sample we perform a jackknife analysis by fitting the density profile of 100 random selected subsamples and taking only $80\%$ of the total lens systems. We find that the distributions of $\sigma_{V}$ and $R_{200}$ follow gaussians with dispersions   $\lesssim 8\%$ , which are considered in their uncertainty estimates.

\section{LENSING MASS DETERMINATIONS}
\label{sec:results}

\subsection{MaxBCG results}

\begin{figure}
\centering
\includegraphics[width=0.48\textwidth]{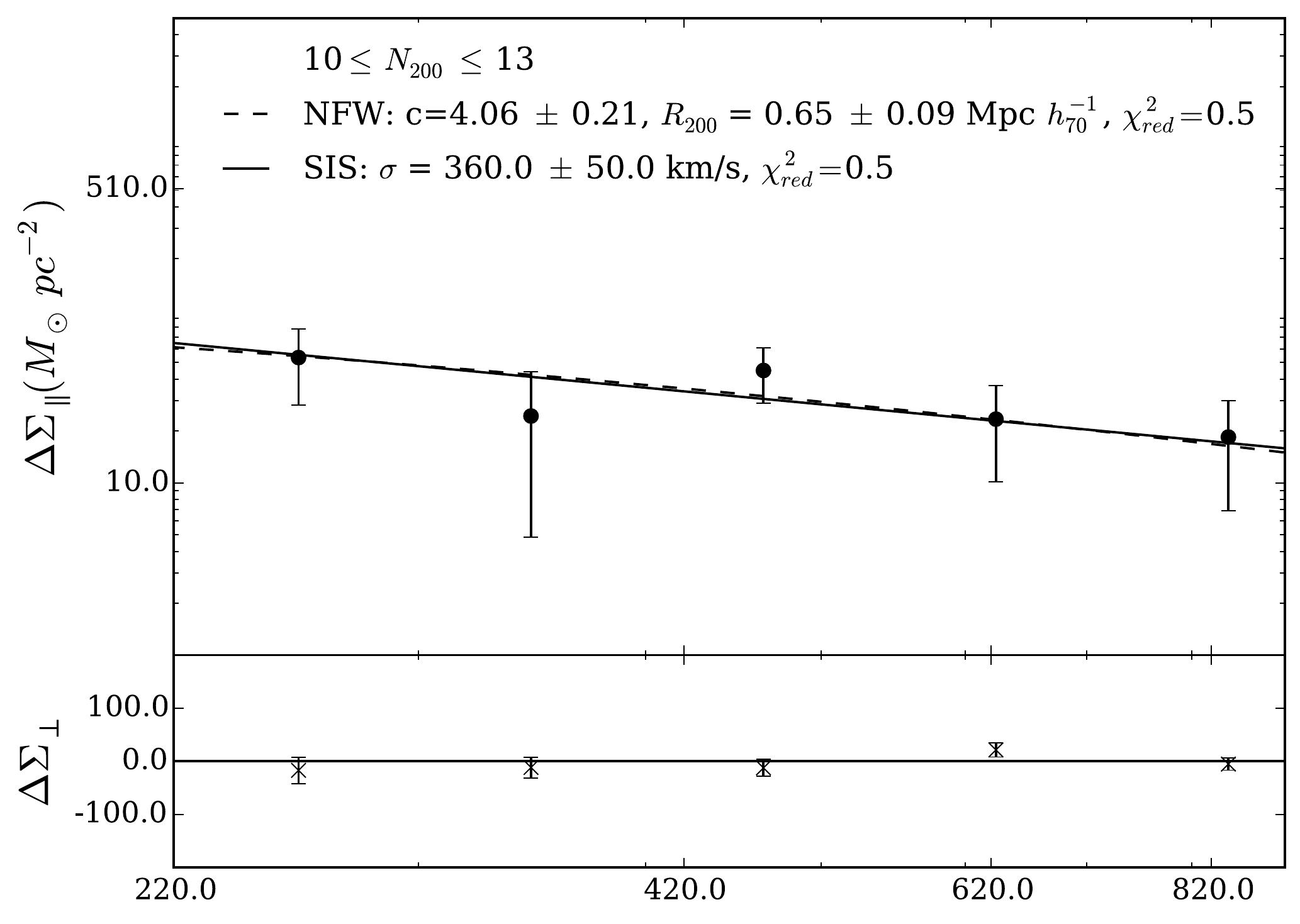}
\includegraphics[width=0.48\textwidth]{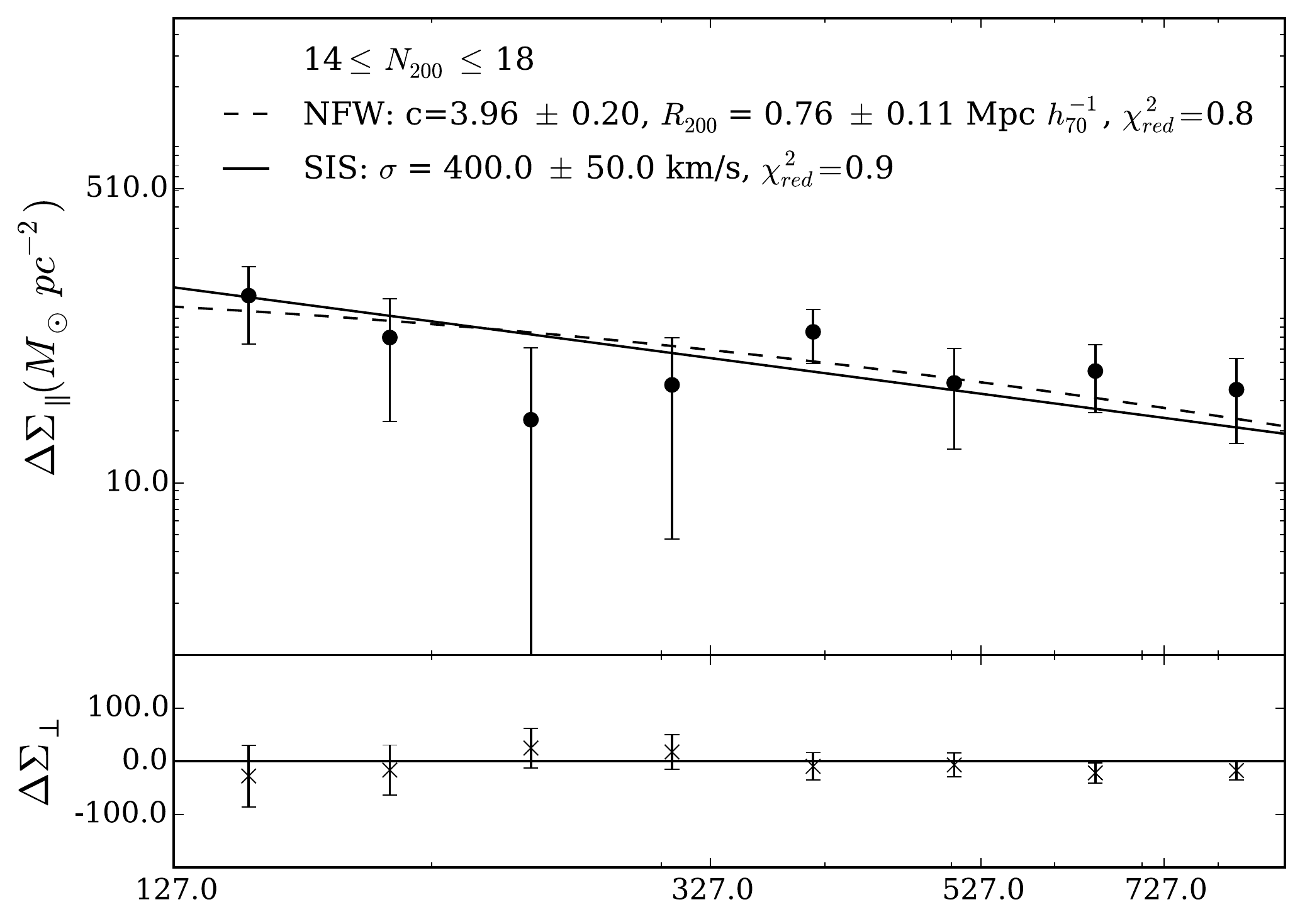}
\includegraphics[width=0.495\textwidth]{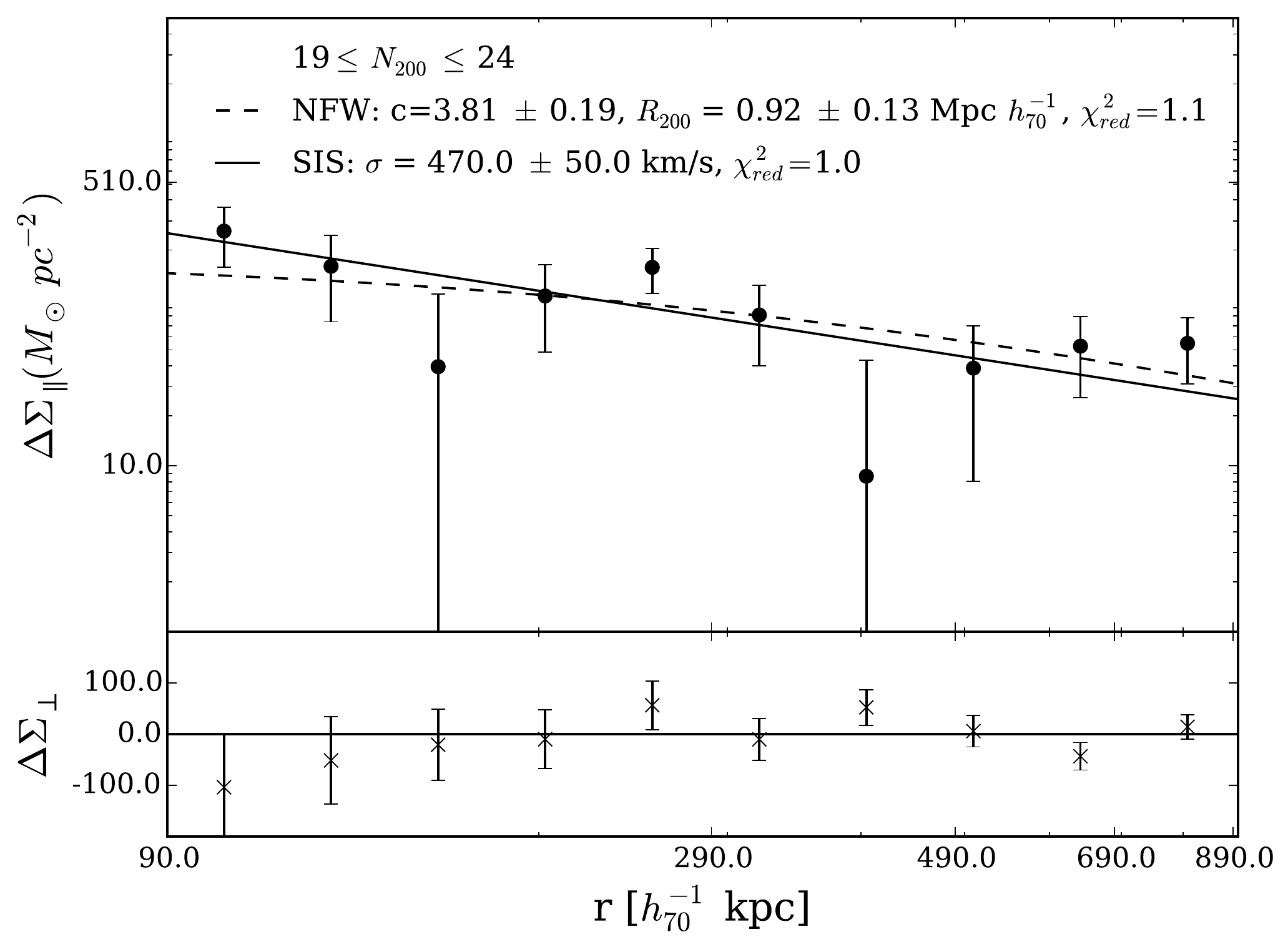}
\caption{Average density contrast $\Delta \Sigma (r)$ profile of maxBCG sample, for different richness bins. The solid and the dashed lines represent the best fit of SIS and NFW profiles, respectively. The lower panels shows the profile resulting of averaging the cross ellipticity component, and should be equals to zero. Error bars are computed according to Equation\,\ref{eq:err}. Derived fitted parameters and errors take into account the discussion of subsection\,\ref{susec:syserrores}}. 
\label{profile_maxbcg}
\end{figure}

We select galaxy systems with $z < 0.25$ from the maxBCG sample since our photometric cuts do not allow us to extend our sample to larger redshifts. Moreover, we only analyse galaxy systems with $N_{200} < 24$, leading to a total sample of 7797 objects. We do not extend our analysis to richer systems given that the low number of clusters with $N_{200} > 24$ (1129) does not allow for a detailed binning. After applying the seeing criterion described previously, the final sample comprises 6701 systems.   

From the stacking analysis we obtain the projected mass density profiles for three $N_{200}$ richness bins (Figure\,\ref{profile_maxbcg}) whose lensing best-fitting parameters are given in Table\,\ref{table:1}. For the richest clusters ($19\leq\ N_{200} \leq24$) the NFW mass is larger by $\sim 20 \%$  than the SIS mass \citep[consistent with previous works, ][]{Gonzalez15,Okabe10}. This could be due to the  shortcoming of the SIS model to fit the curvature of the distortion profile of a NFW halo at large radii \citep{Okabe10}. The sharp fall of the SIS profile on large scales is compensated, but not entirely, by the overestimation of the mass at small radii, which causes an overall mass underestimation.

These results could be easily compared to \citet{Sheldon09} and \citet{Johnston07}. They presented a complete analysis of the whole maxBCG sample extended to $N_{200} = 3$. The total sample includes $\sim 130\,000$ galaxy systems with redshifts ranging from 0.1 to 0.3. For the analysis they selected background galaxies according to individual photometric redshifts. To estimate the masses they modelled the density profile from 25\,$h_{71}^{-1}$kpc up to $\sim 30\,h_{71}^{-1}$\,Mpc, taking into account the BCG halo and neighboring mass concentrations together with the dark matter halo of the galaxy system. They also included corrections regarding miscentering distributions. \citet{Johnston07} provides $M_{200}$ masses for different richness bins. 

\begin{figure}
\centering
\includegraphics[scale=0.31]{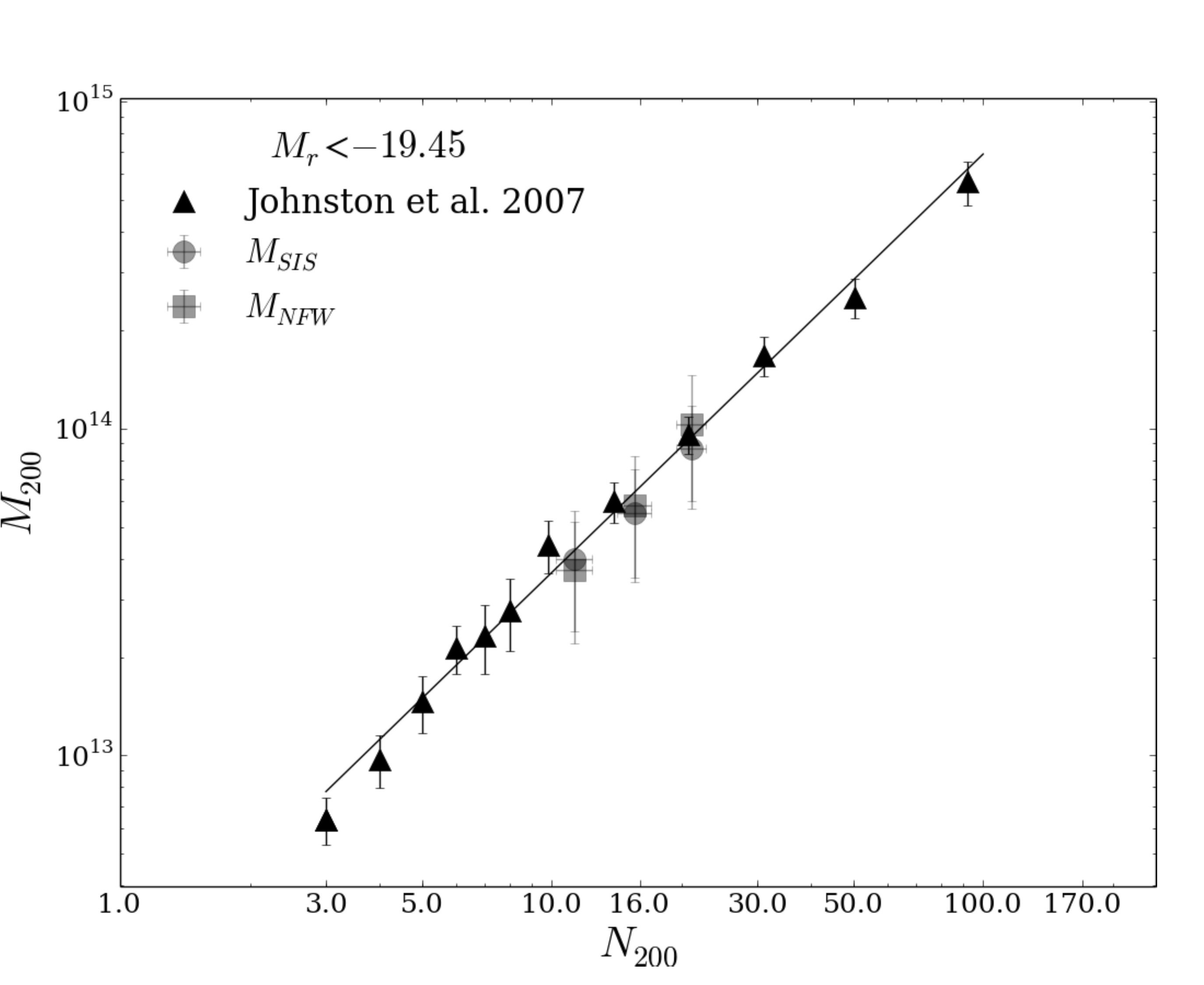}
\caption{Mass-richness relation by \citet{Johnston07} (\textit{solid line}), together with $M_{200}$ masses used to obtain that relation (\textit{triangles}) and our mass estimates  (\textit{squares} for NFW masses and \textit{circles} for SIS masses) vs $N_{200}$ from maxBCG catalogue.}
\label{MN_maxbcg}
\end{figure}

In Figure\,\ref{MN_maxbcg} we compare our lensing mass estimates to the results of \citet{Johnston07}. In spite of our much simpler analysis, it can be seen a good agreement, demonstrating the reliability of our method. As expected, NFW masses have a better correspondence to the mass-richness relation, since this was the model used by \citet{Johnston07} in order to describe the halo component.

\begin{table*}

\caption{MaxBCG results}\label{tab:esp}
\label{table:1}

\begin{tabular}{@{}crrccccrrcr@{}}
\hline
\hline
\rule{0pt}{1.05em}%
  Selection criteria & $\langle N_{200} \rangle$  &   $N_{Lens}$ & $\langle z_{Lens} \rangle$ &$S/N$ & \multicolumn{2}{c}{SIS} & \multicolumn{3}{c}{NFW}   \\
 &   &    & &  &$\sigma_{V}$ & $M_{200}$ & $c_{200}$ & $R_{200}$ & $M_{200}$   \\
 &   &    & &  & [km\,s$^{-1}$] &  [$10^{12} h_{70}^{-1} M_{\odot} $] & & [$h_{70}^{-1}$\,Mpc] & [$10^{12} h_{70}^{-1} M_{\odot} $] \\

 \hline
\rule{0pt}{1.05em}%
$10 \leq N_{200}\leq 13$ & $11.29   \pm 0.02$  &   3854  & $0.18$ & 4.4 &    $360 \pm 50$   &    $40 \pm 16$    & $4.06 \pm 0.21$  & $0.65 \pm 0.09$ &   $37 \pm 15$ \\ 
$14 \leq N_{200}\leq 18$ & $15.64   \pm 0.03$  &   1852  & $0.18$ & 5.4 &    $400 \pm 50$   &    $55 \pm 20$    &   $3.96 \pm 0.20$  & $0.76 \pm 0.11$ &   $58 \pm 24$ \\ 
$19 \leq N_{200}\leq 24$ & $21.15   \pm 0.01$  &   995  & $0.17$ & 6.3 &    $470 \pm 50$   &    $87 \pm 30$    &   $3.81 \pm 0.19$  & $0.92 \pm 0.13$ &   $103 \pm 43$ \\

\hline         
\end{tabular}
\medskip
\begin{flushleft}
\textbf{Notes.} Columns: (1) $N_{200}$ bins; (2) mean $N_{200}$ and the standard deviation of the mean; ; (3) number of groups considered in the stack; (4) average $z$ of the considered samples; (5) $S/N$ ratio as defined in Equation\,\ref{SN}; (6) and (7) results from the SIS profile fit, velocity dispersion and $M^{SIS}_{200}$; (8), (9) and (10), results from the NFW profile fit, $c_{200}$ adopted according $M^{SIS}_{200}$ and $\langle z_{Lens} \rangle$ (see text for details), $R_{200}$ and $M^{NFW}_{200}$. 
\end{flushleft}
\end{table*}

\subsection{Yang groups results}
\label{subsec:resone}

\begin{figure*}
\centering
\includegraphics[scale=0.4035]{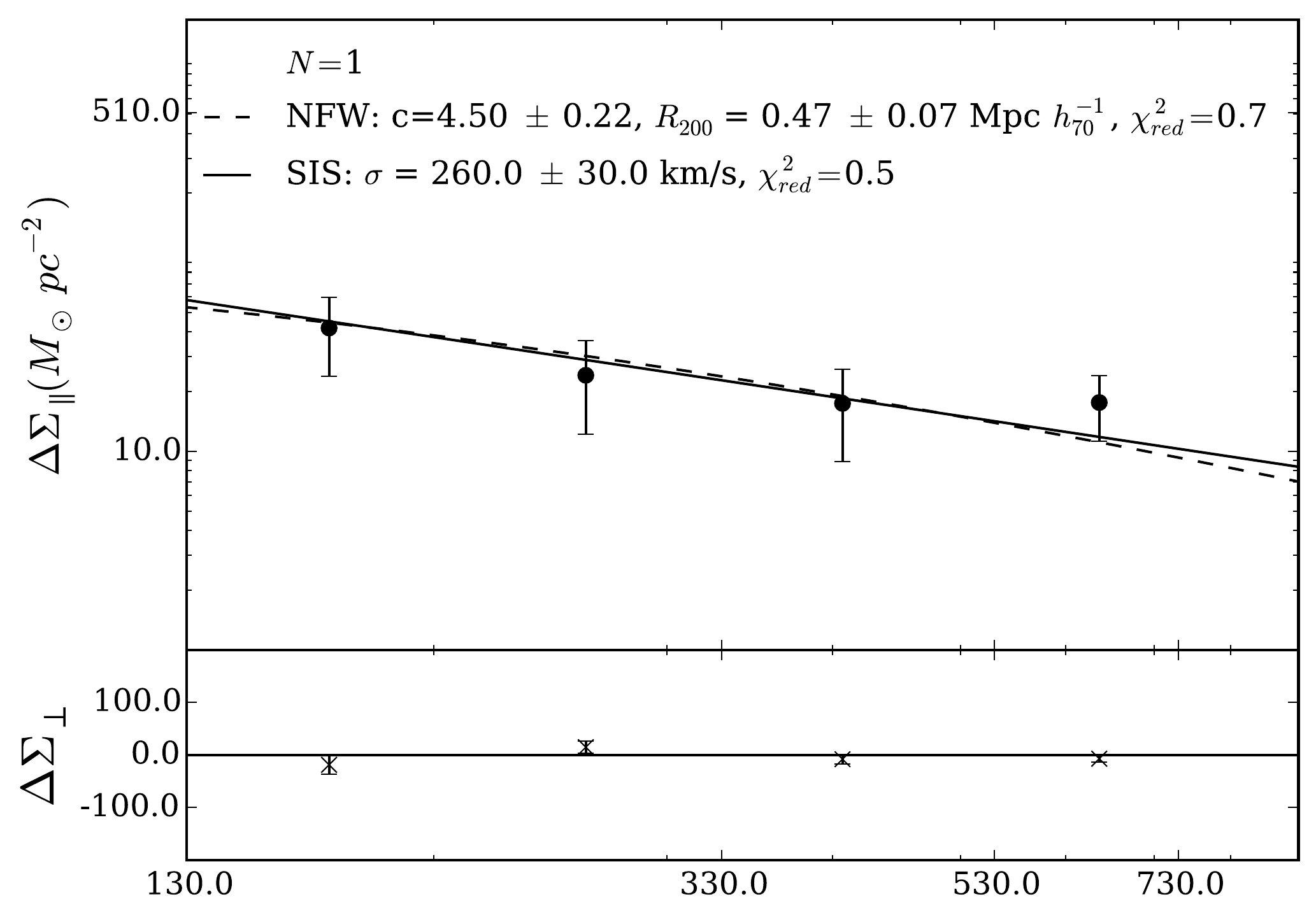}
\includegraphics[scale=0.385]{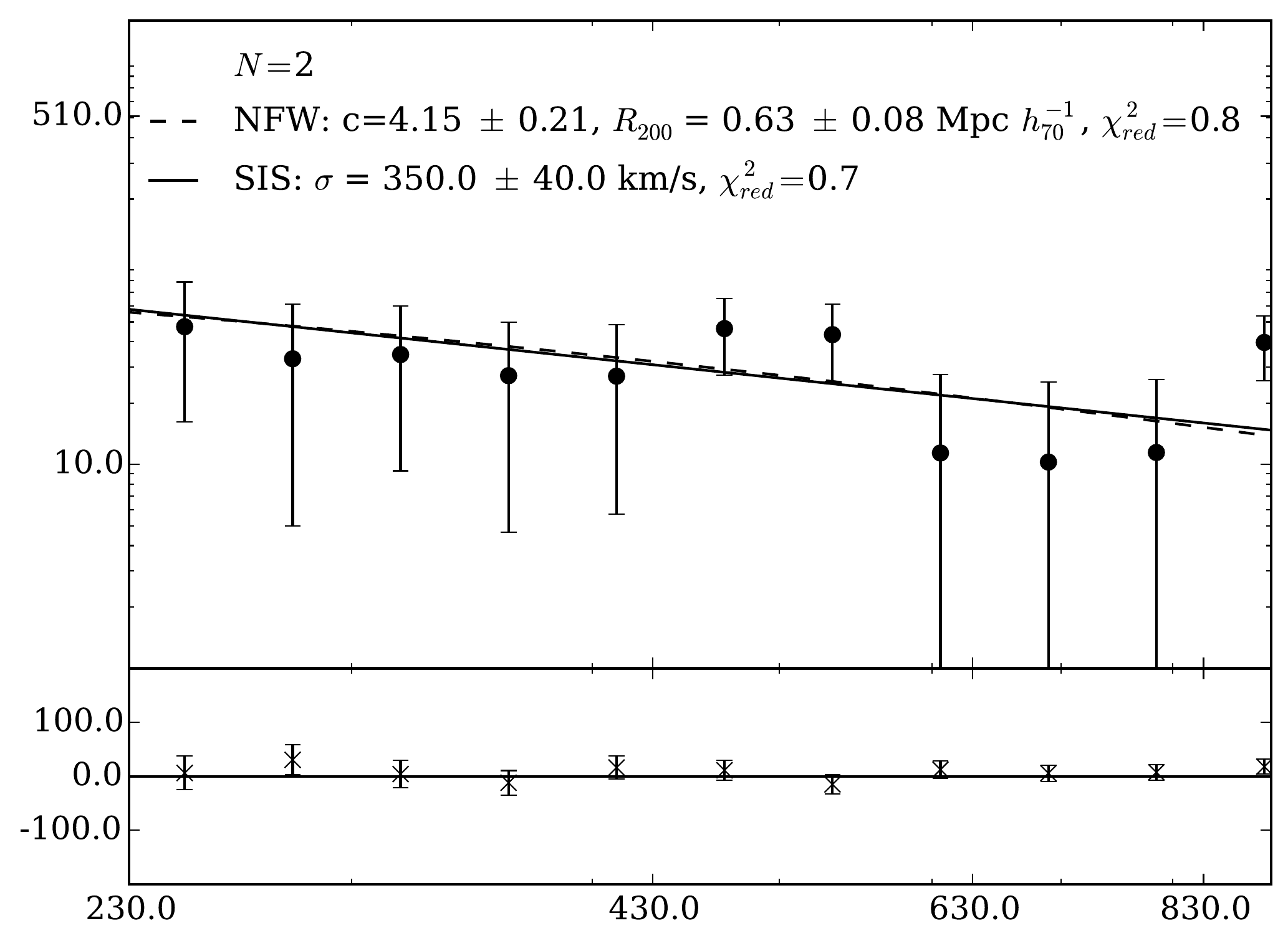}\\
\includegraphics[scale=0.4035]{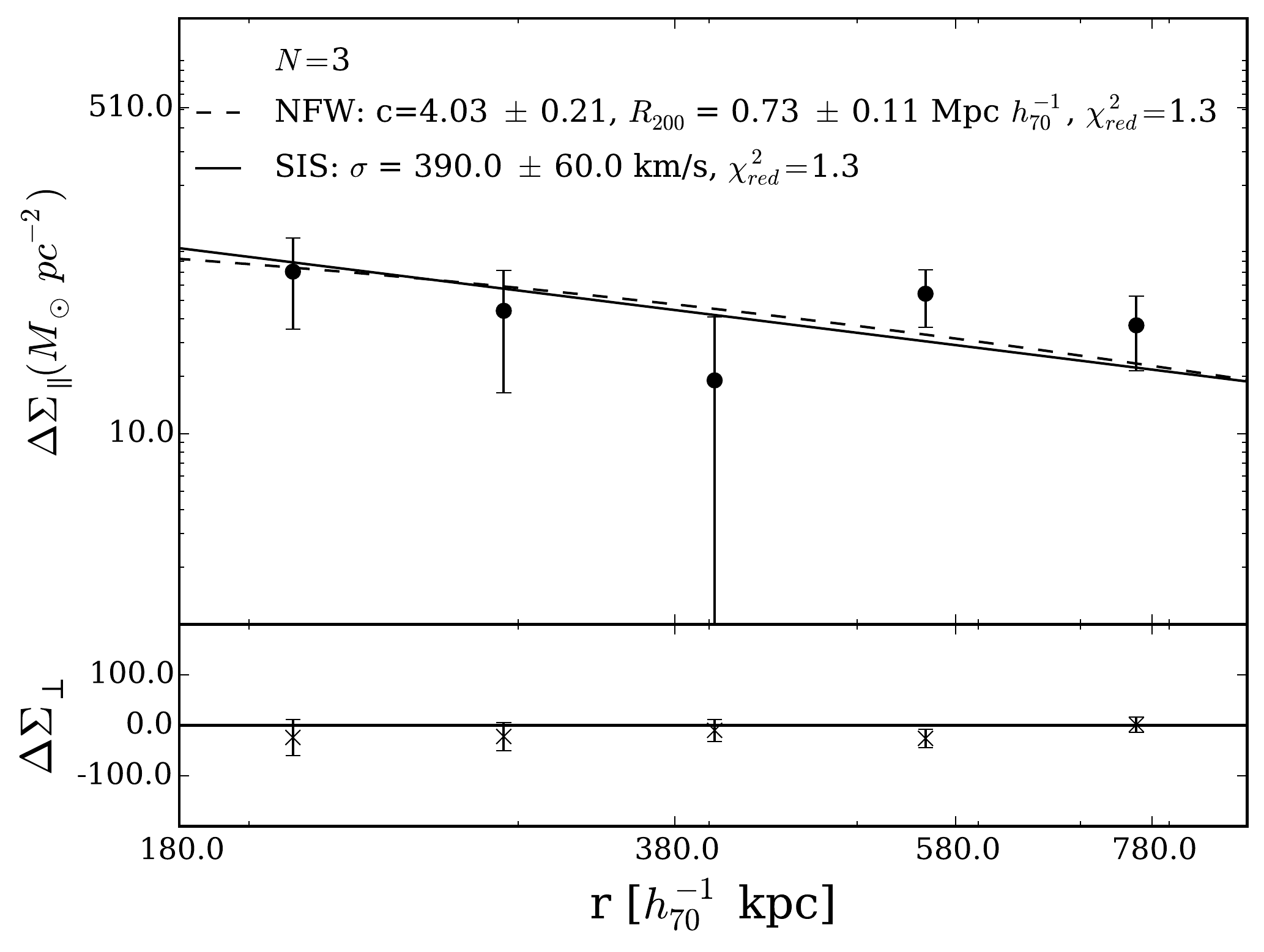}
\includegraphics[scale=0.385]{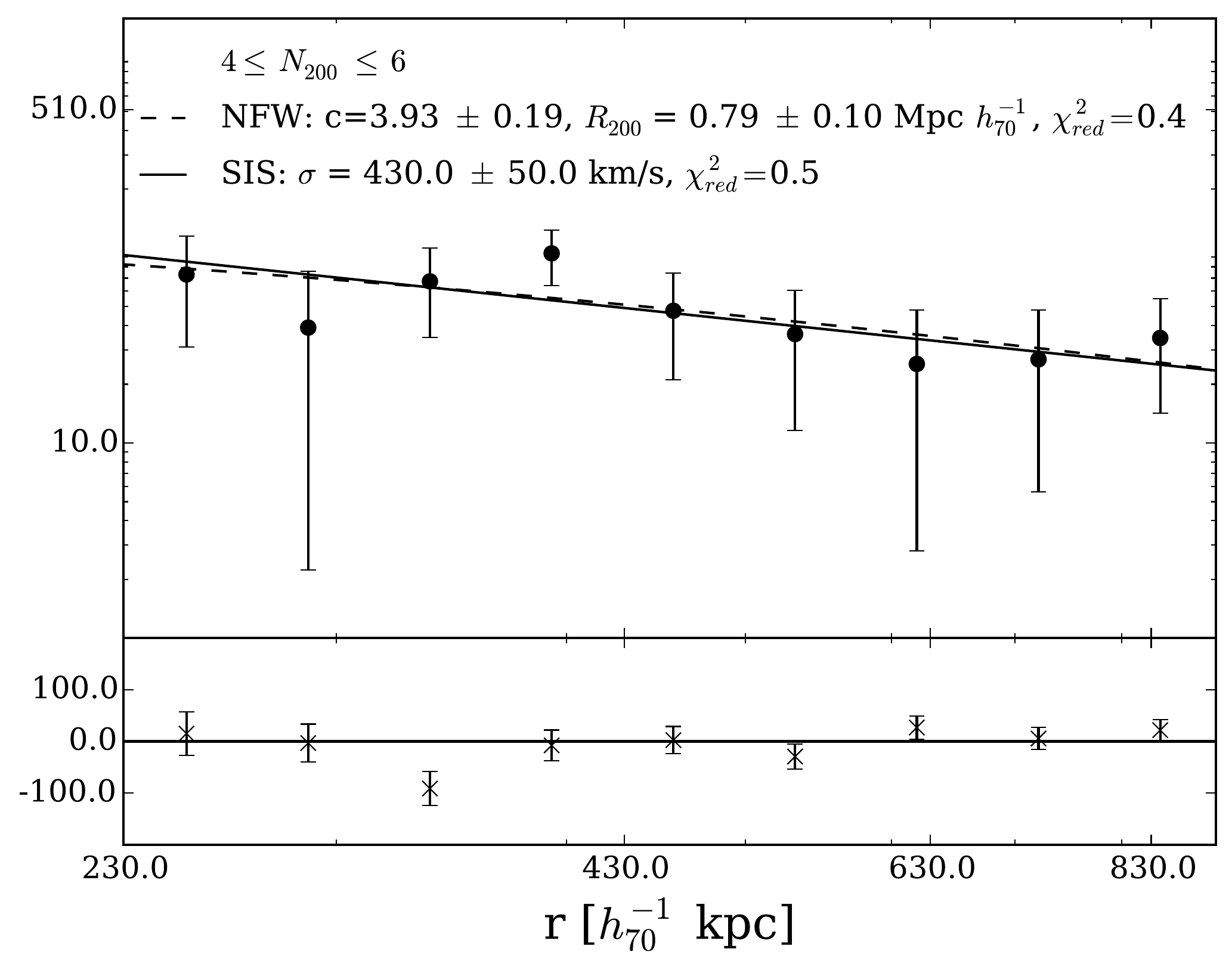}
\caption{Same as Figure\,\ref{profile_maxbcg}, for the Yang sample.}
\label{profile_yang}
\end{figure*}

\begin{table*}

\caption{Yang groups results}\label{tab:esp}
\label{table:2}

\begin{tabular}{@{}crrccccrrcr@{}}
\hline
\hline
\rule{0pt}{1.05em}%
  Selection criteria  &   $N_{Lens}$  &$\langle z_{Lens} \rangle$ &$S/N$ & \multicolumn{2}{c}{SIS} & \multicolumn{3}{c}{NFW}   \\
 &     & &  &$\sigma_{V}$ & $M_{200}$ & $c_{200}$ & $R_{200}$ & $M_{200}$   \\
   &    & &  & [km\,s$^{-1}$] &  [$10^{12} h_{70}^{-1} M_{\odot} $] & & [$h_{70}^{-1}$\,Mpc] & [$10^{12} h_{70}^{-1} M_{\odot} $] \\
 \hline
\rule{0pt}{1.05em}%
$N = 1$    &   7348  & $0.13$  & 4.6 &    $260 \pm 30$   &    $15 \pm 6$    & $4.50 \pm 0.22$  & $0.47 \pm 0.07$ &   $13 \pm 5$ \\
$N = 2$    &   4875  & $0.15$  & 5.5 &    $350 \pm 40$   &    $35 \pm 13$    & $4.15 \pm 0.21$  & $0.63 \pm 0.08$ &   $32 \pm 13$ \\
$N = 3$    &  1669  & $0.14$  & 4.9 &    $390 \pm 60$   &    $52 \pm 22$    & $4.03 \pm 0.21$  & $0.73 \pm 0.11$ &   $50 \pm 22$ \\
$4 \leq N \leq 6$    &  1698  & $0.14$  & 5.4 &    $430 \pm 50$   &    $71 \pm 26$    & $3.93 \pm 0.19$  & $0.79 \pm 0.10$ &   $64 \pm 25$ \\

\hline         
\end{tabular}
\medskip
\begin{flushleft}
\textbf{Notes.} Columns: (1) Selection criteria to limit the sample of groups for stacking; (2) number of groups considered in the stack; (3) average $z$ of the sample; (4) $S/N$ ratio as defined in Equation\,\ref{SN}; (5) and (6) results from the SIS profile fit, the velocity dispersion and $M^{SIS}_{200}$; (7), (8) and (9), results from the NFW profile fit, $c_{200}$ adopted according $M^{SIS}_{200}$ and $\langle z_{Lens} \rangle$ (see text for details), $R_{200}$ and $M^{NFW}_{200}$. 
\end{flushleft}
\end{table*}

\begin{figure*}
\begin{multicols}{2}

\includegraphics[scale=0.42]{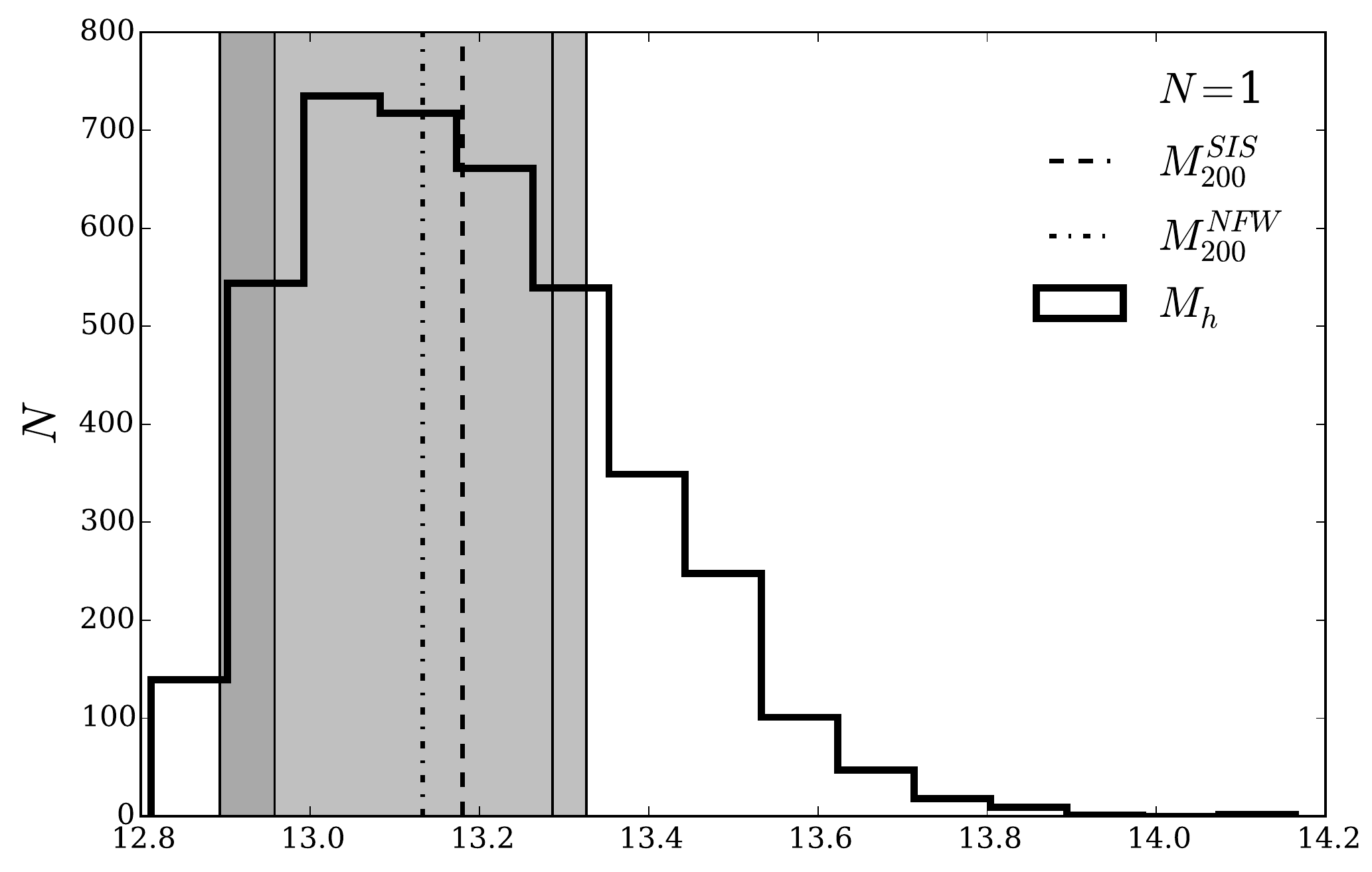}
\includegraphics[scale=0.42]{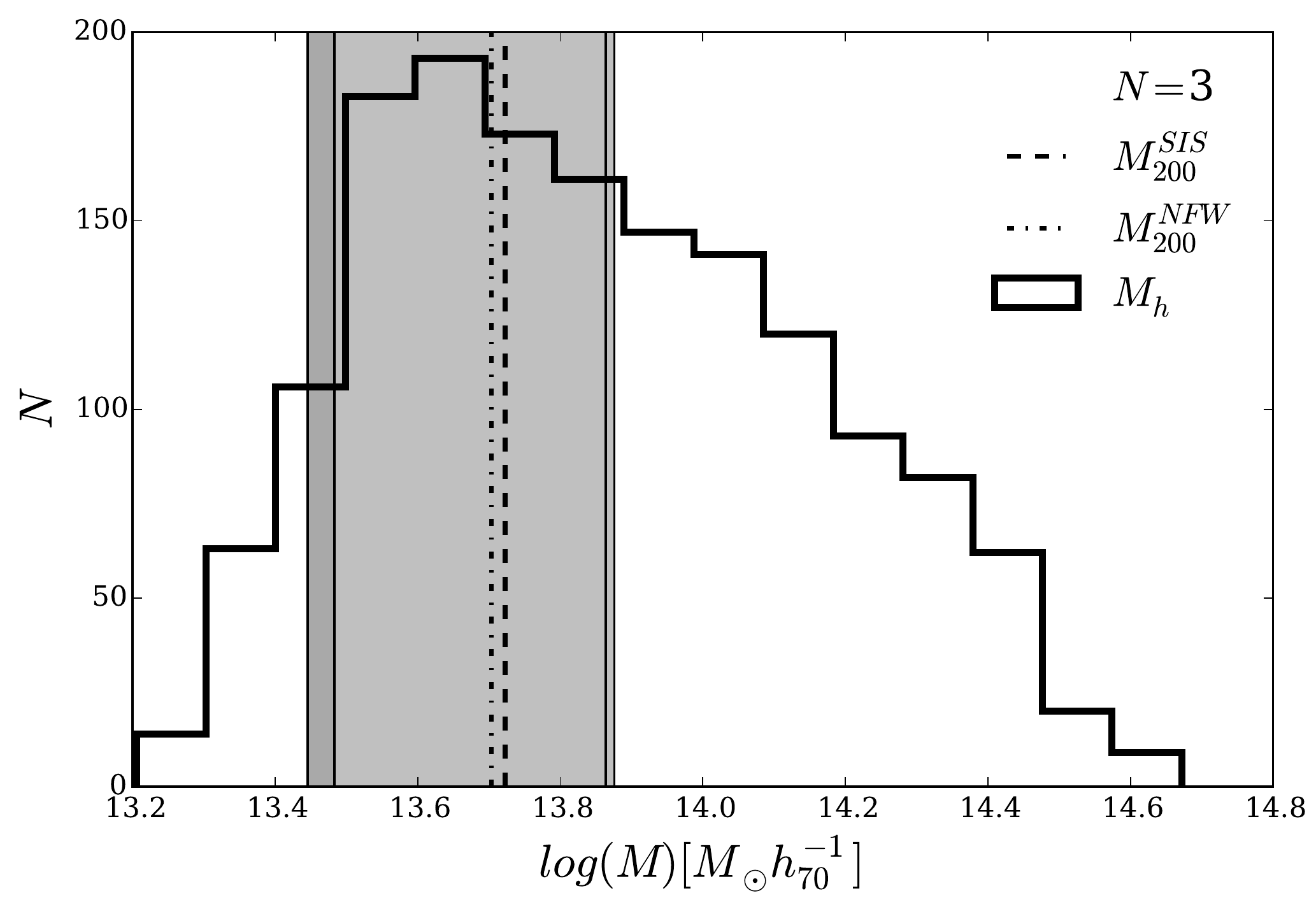}
\includegraphics[scale=0.395]{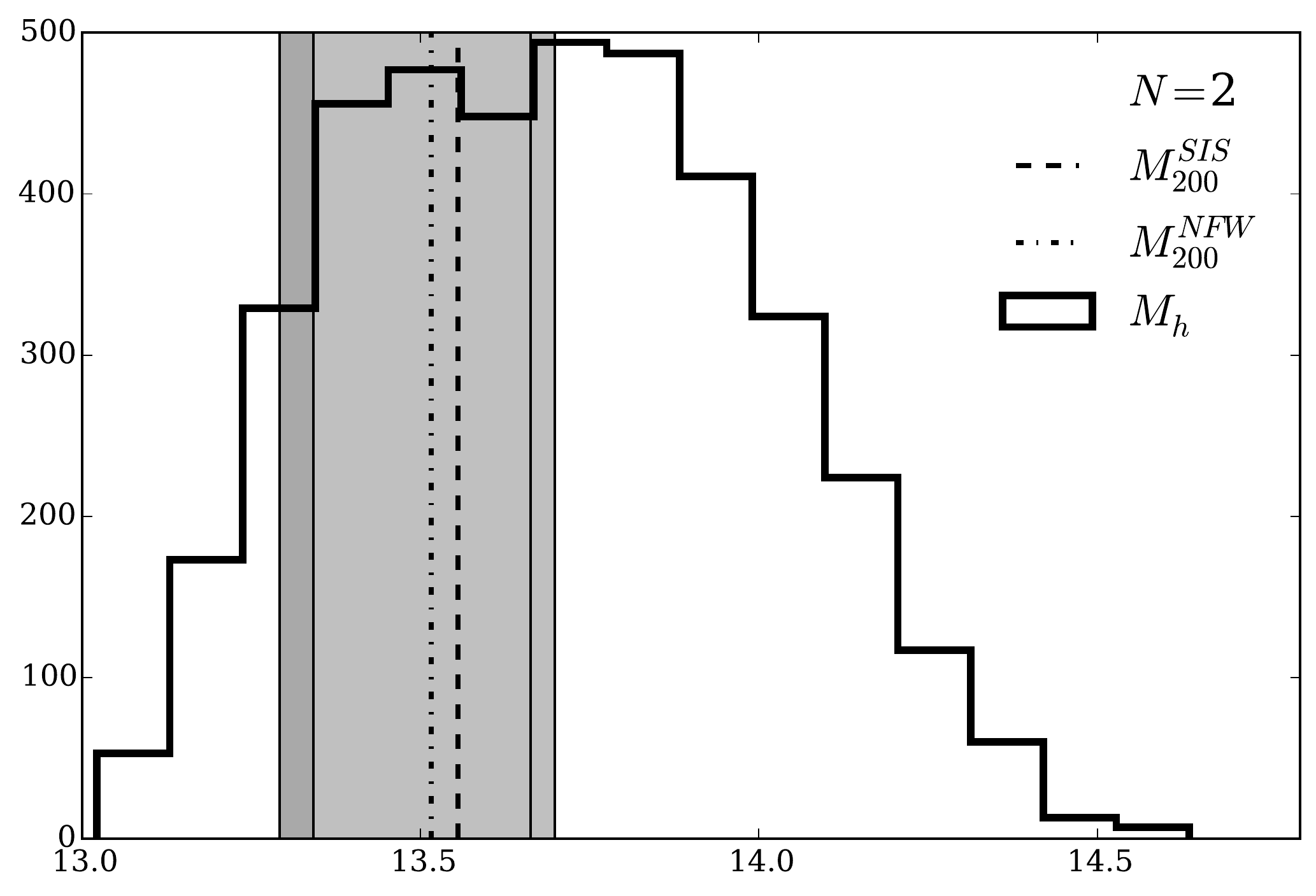}
\includegraphics[scale=0.406]{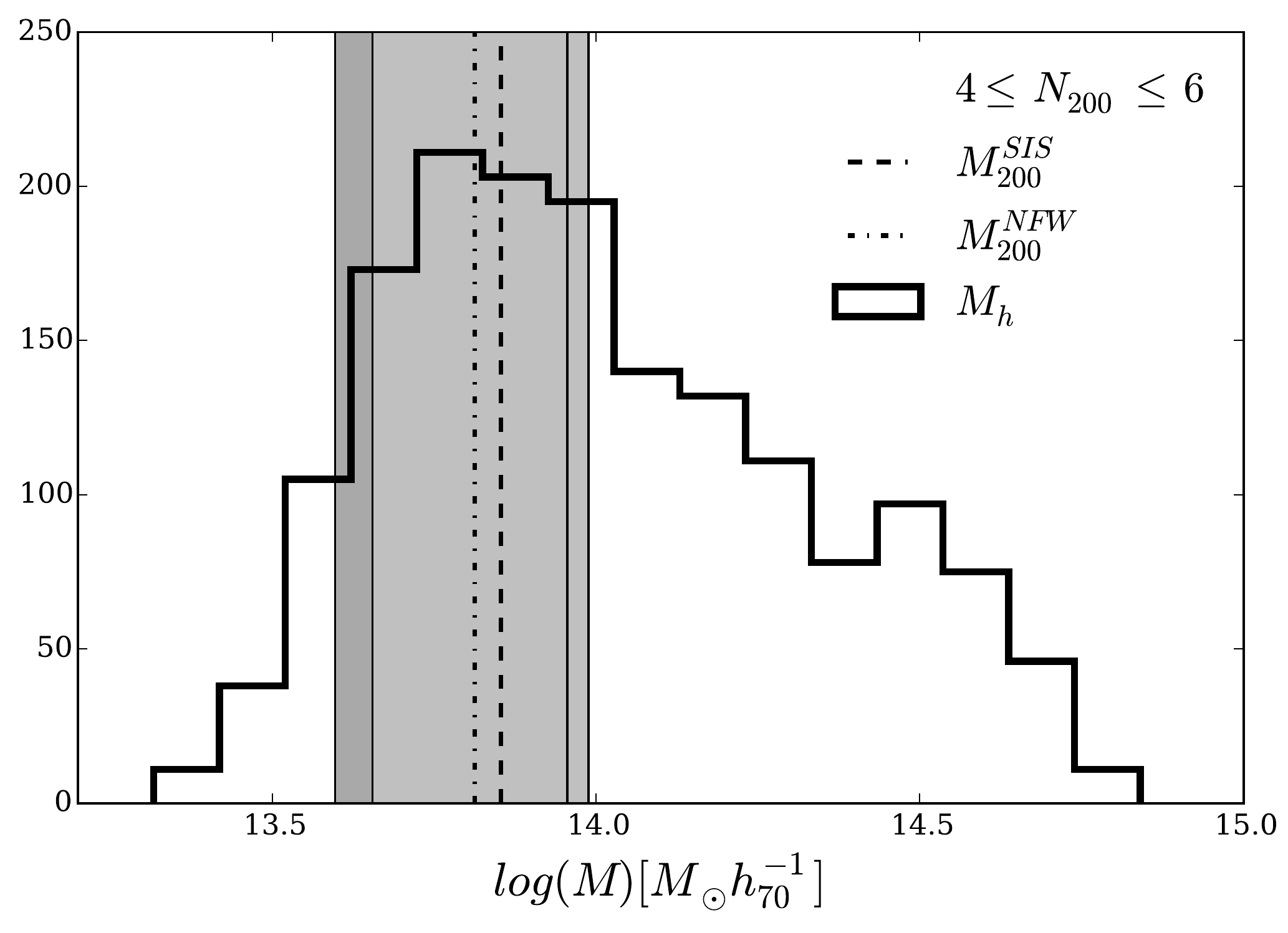}

\end{multicols}
\caption{Distribution of halo masses computed by \citet{Yang12} in the four richness bins, together with our lensing masses estimates (vertical lines and shaded regions for the corresponding errors)}
\label{fig:masas}
\end{figure*}

We determine the mean mass for four subsamples of the Yang group catalogue described in subsection\,\ref{yang_sample}: $N_{members} = 1, 2, 3$ and 4 to 6. As it was done previously for the maxBCG sample, only groups in $i-$band frames with seeing values lower than 1.3" are analysed. The density profiles are shown in Figure\,\ref{profile_yang} and their best-fit parameters are given in Table\,\ref{table:2}. NFW and SIS masses are in good agreement, $\langle M^{NFW}_{200}/M^{SIS}_{200} \rangle = 0.91 \pm 0.03$, showing that in contrast to massive clusters, a SIS profile is a suitable model to describe the mass distribution of low mass systems.

In Figure\,\ref{fig:masas} we plot the distribution of $M_h$ obtained by \citet{Yang12} for each subsample together with our lens mass determinations. For the four subsamples we observe a good agreement between our lens masses and masses derived from mass-to-light ratios.

\begin{figure*}
\centering
\includegraphics[scale=0.35]{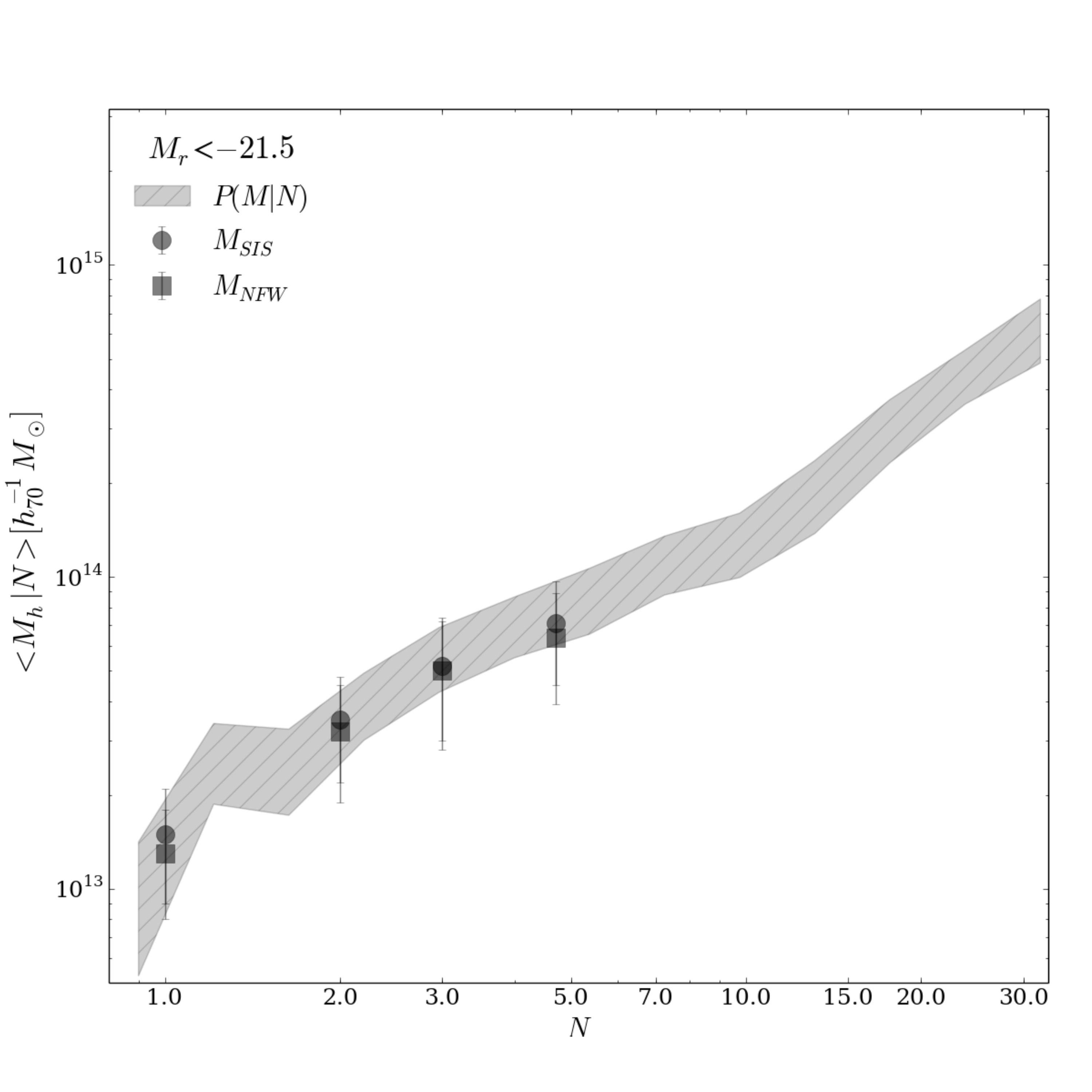}
\caption{$P (M \mid N)$ obtained for SDSS DR7 implementing background subtraction method for a limiting absolute magnitude $M^{lim}_{r} = -21.5$. Squares and circles represent weak lensing $M^{NFW}_{200}$ and $M^{SIS}_{200}$ masses versus $N$, respectively.}
\label{HOD_yang}
\end{figure*}

We use these results to compare them with the  $P (M \mid N)$ relation. As explained in Section\,\ref{sec:theory}, HOD cannot be directly compared to the lensing mass-richness relation, so it is necessary to compute $P (M \mid N)$. We derive this distribution by using the same background subtraction method as in \citet{Rodriguez15}, computing  average halo masses in richness bins. We consider only galaxy group members with $M_{r} < -21.5$. Hence, these distribution can be directly compared to the mass-richness relation obtained from this sample of groups. In Figure\,\ref{HOD_yang} we plot $P (M \mid N)$ and $M_{200}$ vs $N$. As it can be noticed that lens mass determinations by both models, SIS and NFW, agree with the $P (M \mid N)$ relation.

\section{DISCUSSION AND CONCLUSIONS}
\label{sec:test}
In this work we derive a mass-richness relation obtained by a weak lensing analysis and compare it with $P(M \mid N)$ estimated through a straightforward background subtraction technique. To test our lens analysis, we estimated masses for a sample of maxBCG clusters by modeling the dark matter halo using both, SIS and NFW model mass profiles. Our results are consistent with the mass-richness relation obtained by \citet{Johnston07}, who analysed an extended maxBCG sample with richness values $N_{200} \geq 3$. We have also performed a weak lensing analysis for a sample of low richness groups from the catalogue of \citet{Yang12} with results well described by NFW and SIS models.

Following the technique described by \citet{Rodriguez15}, we computed $P(M \mid N)$ from the same data set used to estimate HOD restricted to a limiting absolute magnitude $M_{r} = -21.5$. This distribution is compared to the mass-richness relation obtained for the Yang group sample in Figure\,\ref{HOD_yang} where it can be seen a good agreement between both relations.  In particular we stress the fact that our result for  $N_{member} = 1$ is in consistent with $P (M \mid N)$. These groups include systems that are composed by only one galaxy with redshift information available, making it impossible to estimate their virial masses. However, our stacking lensing analysis allows to derive the average system mass.

The agreement between $M-N$ relation and $P(M \mid N)$ reinforces the confidence in the method employed in computing HOD based in background subtraction techniques. Besides, it presents a new approach to test the mass-richness relation. It is important to highlight that these results can not be directly compared to other HOD analysis \citep[eg. ][]{Tinker12, Guo15} since we adopt richness instead of mass bins.
In our analysis we used the same information as in the computation of HOD to obtain $P (M \mid N)$ in a straightforward way, allowing for a direct comparison between two independent relations, lensing mass-richness and $P (M \mid N)$. This result can be extended to fainter limiting magnitudes which could provide a deeper understanding of the relation between galaxies and mass distribution in halos.

\section*{Acknowledgements}

We thank to Miriam Zorn for her careful reading and corrections to the manuscript.

We thank the anonymous referee for their very useful comments that improved the content and clarity of the manuscript.
This work was partially supported by the Consejo Nacional de Investigaciones Cient\'{\i}ficas y T\'ecnicas (CONICET, Argentina) 
and the Secretar\'{\i}a de Ciencia y Tecnolog\'{\i}a de la Universidad Nacional de C\'ordoba (SeCyT-UNC, Argentina).
We made an extensive use of the following python libraries:  http://www.numpy.org/, http://www.scipy.org/, http://roban.github.com/CosmoloPy/ and http://www.matplotlib.org/.\\

\bibliographystyle{mn2e}
\bibliography{references}

\appendix

\end{document}